\documentclass[aps,prl,twocolumn,showpacs,amsmath,amssymbsuperscriptaddress]{revtex4-1}

\usepackage{amsmath,amsfonts,bm,graphicx,verbatim,mathrsfs,units}
\usepackage[colorlinks=true,citecolor=blue]{hyperref}
\usepackage{pdfpages}

\newcommand{\ket}[1]{\left| #1 \right \rangle}
\newcommand{\bra}[1]{\left \langle #1 \right|}

\begin{document}
\title{Floquet Theory of Electron Waiting Times in Quantum-Coherent Conductors}
\date{\today}

\author{David Dasenbrook}
\author{Christian Flindt}
\author{Markus B\"uttiker}
\affiliation{D\'epartement de Physique Th\'eorique, Universit\'e de Gen\`eve, 1211
  Gen\`eve, Switzerland}

\begin{abstract}
  We present a Floquet scattering theory of electron waiting time distributions in periodically driven quantum conductors. We employ a second-quantized formulation that
  allows us to relate the waiting time distribution to the Floquet scattering matrix of the system. As an application we evaluate the electron waiting times for a quantum point contact, modulating either the applied voltage (external driving) or the transmission probability (internal driving) periodically in time. Lorentzian-shaped voltage pulses are of particular interest as they lead to the emission of clean single-particle excitations as recently demonstrated experimentally. The distributions of waiting times provide us with a detailed characterization of the dynamical properties of the quantum-coherent conductor in addition to what can be obtained from the shot noise or the~full~counting~statistics.
\end{abstract}

\maketitle

\paragraph{Introduction.---}
A surge of interest in dynamic quantum conductors has recently led to a number of ground-breaking experiments \cite{gabelli06,feve07,bocquillon13,dubois13nature,flindt13,fletcher13}. An on-demand coherent single-electron source based on a submicron capacitor \cite{gabelli06,buttiker93} has been experimentally realized and successfully operated in the gigahertz regime \cite{feve07}. Recently, the fermionic analogue of an optical Hong-Ou-Mandel experiment was performed to demonstrate that two such on-demand sources produce indistinguishable electronic quantum states \cite{bocquillon13}. Additionally, clean single-particle excitations have been created on top of a Fermi sea by applying a periodic sequence of Lorentzian-shaped voltage pulses to an electrical contact \cite{dubois13nature,flindt13} following a pioneering theoretical proposal by Levitov and co-workers \cite{levitov96,ivanov97,keeling06}.

These experimental breakthroughs hold promises for future gigahertz quantum electronics with precisely synchronized single-particle operations. One may envision circuit architectures with driven single-electron emitters coupled to the edge states of a quantum Hall conductor (or to the helical edge states in a topological insulator~\cite{hofer13,inhofer13}) serving as rail tracks for charge and information carriers by guiding them to beam splitters (quantum point contacts) and particle interferometers for further processing. To facilitate progress towards this goal, a detailed understanding of the single-particle emitters and their statistical properties is required.

In one approach, the full counting statistics of emitted charge is analyzed \cite{Blanter00,Nazarov03,andreev00,makhlin01,albert10}. The charge fluctuations are
typically integrated over many periods of the driving and important short-time physics may be lost. In a complementary approach, one considers the distribution of waiting times between charge carriers \cite{brandes08,albert11,albert12,thomas13,rajabi13,albert14,thomas14,tang14}. This view on quantum transport seems promising as picosecond single-electron detection is now becoming feasible \cite{fletcher13}. A quantum theory of electron waiting times has recently been developed for voltage-biased mesoscopic conductors \cite{albert12}, however, so far without an explicit driving. To describe the statistical properties of coherent single-electron emitters, a theory of waiting time distributions (WTD) for driven mesoscopic conductors is clearly desirable.

\begin{figure}
 \includegraphics[width=0.95\columnwidth]{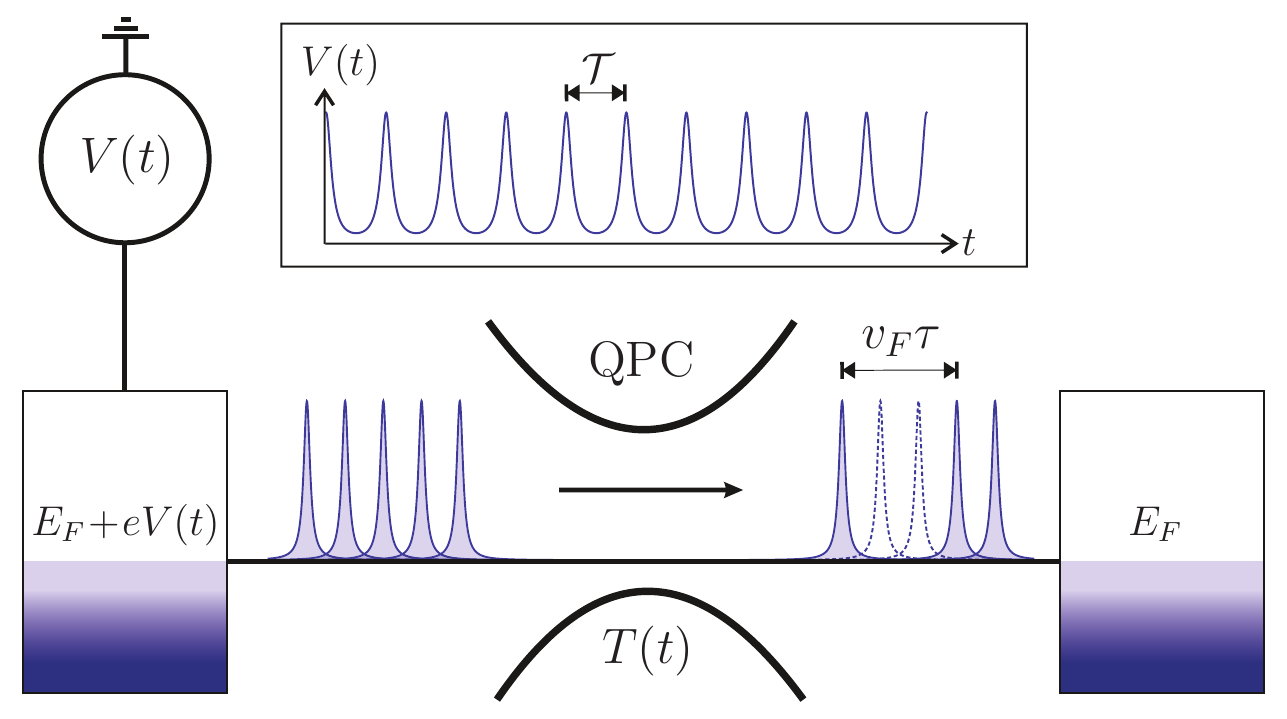}
  \caption{(color online). Driven quantum point contact. Lorentzian-shaped voltage pulses $V(t)$ with period $\mathcal{T}$ generate a train of clean single-electron excitations (levitons) above the Fermi level $E_F$. The levitons propagate with the Fermi velocity $v_F$ towards a quantum point contact (QPC) whose transmission $T(t)$ can be controlled. We are interested in the distribution of waiting times $\tau$ between transmitted electrons. Reflected (missing) levitons are indicated by dashed lines.}
  \label{fig:levitonwtdsetup}
\end{figure}

In this Letter we develop a quantum formalism for electron waiting times in dynamic mesoscopic conductors described by Floquet scattering theory \cite{moskaletsbook11,pederson98}. Our methodology is applicable to a wide range of periodically driven mesoscopic conductors, for instance the quantum point contact (QPC) depicted in Fig.~\ref{fig:levitonwtdsetup}. We illustrate our method by evaluating the WTDs under two different driving schemes.
Lorentzian-shaped voltage pulses applied to the QPC \cite{levitov96,ivanov97,keeling06} produce clean single-electron excitations (levitons)
as recently demonstrated experimentally \cite{dubois13nature,flindt13}. We analyze the distribution of waiting times between levitons transmitted through the QPC.
We then fix the voltage and investigate the waiting time between electrons with the transmission $T(t)$ of the QPC modulated periodically in time. We focus here on electronic conductors, but our ideas may also be realized in cold atomic gases~\cite{brantut12}.

\paragraph{Formalism.---}
We consider a central scatterer connected to electronic leads. We are interested in the distribution $\mathcal{W}(\tau)$ of waiting times $\tau$ between electrons scattered from the left to the right lead, passing a particular point $x_0$ in the right lead. A fundamental building block of our theory is the idle time probability (ITP) $\Pi(\tau,t_0)$: The ITP is the probability that \emph{no} transmitted charges are observed at $x_0$ in the time interval $[t_0,t_0+\tau]$. For stationary systems, the ITP is independent of $t_0$ such that $\Pi(\tau,t_0)=\Pi(\tau)$. The WTD can then be expressed as $\mathcal{W}(\tau) = \langle \tau \rangle \partial_\tau^2 \Pi(\tau)$, where $\langle \tau \rangle$ is the mean waiting time~\cite{vyas88,albert12}. In contrast, for the periodically driven systems of interest here, the ITP is a two-time quantity depending on both $t_0$~and~$\tau$. In that case, the WTD can be evaluated by averaging the ITP over a period of the driving $\mathcal{T}$, using $\Pi(\tau)=\int_0^\mathcal{T}\mathrm{d} t_0 \Pi(\tau,t_0)/\mathcal{T}$ above.

Next, we evaluate the ITP for the outgoing many-body state of the scattering problem. Close to the Fermi level we can linearize the dispersion relation around the Fermi energy, $E_k = \hbar v_F k$, such that all electrons scattered into the right lead propagate with the Fermi velocity~$v_F$ towards $x_0$. The probability of finding no transmitted charges at $x_0$ in the temporal interval $[t_0,t_0+\tau]$ is then equal to the probability of finding no transmitted charges in the spatial interval~$[x_0,x_0+v_F\tau]$ at time $t_0+\tau$. We thus define the single-particle projection operator $\widehat{Q}_\tau = \int_{x_0}^{x_0+v_F\tau}\mathrm{d} x\ket{x}\!\bra{x} $ which measures the probability of finding a given particle in the spatial interval~$[x_0,x_0+v_F\tau]$ \cite{hassler08,albert12}. The complementary projector $1-\widehat{Q}_\tau$ similarly measures the probability of $\emph{not}$ finding the particle. To evaluate the ITP for the outgoing many-body state we proceed with a general second-quantized formulation by introducing the operators $\hat{b}^{(\dagger)}_\alpha(E)$ which annihilate (create) electrons in an outgoing state of lead $\alpha=L,R$ at energy $E$. We may then write $\widehat{Q}_\tau = \sum_{E,E^{'}} \int_{x_0}^{x_0+v_F\tau} \mathrm{d} x \varphi^*_{R,E^{'}}(x)\varphi_{R,E}(x) \hat{b}_{R}^\dag(E) \hat{b}_{R}(E^{'})$, where
$\varphi_{R,E}(x)=\bra{x}\hat{b}_{R}^\dag(E)\ket{0}$ and $\ket{0}$ is the vacuum.

The corresponding many-body operator that measures the probability of not finding \emph{any} particles in the spatial interval is the normal-ordered exponential of
$-\widehat{Q}_{\tau}$, see e.~g.~Refs.~\cite{vyas88,levitov96,saito92}. The ITP is then
\begin{equation}
  \Pi(\tau,t_0) = \left \langle : \!e^{ -\widehat{Q}_\tau}\! : \right \rangle_{t_0+\tau},
\label{eq:normalorderedexp}
\end{equation}
with $:\! \dots \!:$ denoting the normal ordering of operators and the expectation value
is taken with respect to the outgoing many-body state evaluated at the time
$t_0+\tau$. Equation~(\ref{eq:normalorderedexp}) is a powerful formal result. It is also
of practical use as it can be applied in a wide range of problems. Below, we consider
noninteracting electrons, but Eq.~(\ref{eq:normalorderedexp}) may equally well form the
basis of a theory of WTDs in interacting systems. For stationary scattering problems,
Eq.~\eqref{eq:normalorderedexp} reduces to the first-quantized result $\Pi(\tau) = \left
  \langle \bigotimes_{i=1}^N [1-\widehat{Q}_\tau] \right \rangle_{t_0+\tau}$ from
Ref.~\cite{albert12} with the expectation value taken with respect to a time-evolved
Slater determinant describing $N$ particles.

\paragraph{Floquet theory.---}
Several recent experiments have realized coherent few-electron emitters operating in the gigahertz regime~\cite{gabelli06,feve07,bocquillon13,dubois13nature,flindt13,fletcher13}. We now use Eq.~\eqref{eq:normalorderedexp} to evaluate the ITP for such mesoscopic scatterers driven with frequency $\Omega = 2\pi/\mathcal{T}$. To this end, Floquet scattering theory provides us with a convenient framework \cite{Moskalets02,moskaletsbook11}. The scatterer is described by the Floquet scattering matrix $\mathcal{S}$ whose matrix elements $\mathcal{S}_{\alpha\beta}(E_n,E)$ with $E_n=E+n\hbar\Omega$ are the amplitudes for an incoming electron in lead $\beta$ with energy $E$ to scatter into lead $\alpha$ having absorbed ($n>0$) or emitted ($n<0$) $|n|$ energy quanta of size $\hbar\Omega$. The operators for the outgoing states can be expressed in terms of the operators of the
incoming states as \cite{Moskalets02,moskaletsbook11}
\begin{equation}
  \label{eq:SFmatrix}
  \hat{b}_{\alpha}(E) = \sum_{\beta}\sum_{E_n} \mathcal{S}_{\alpha\beta}(E,E_n) \hat{a}_\beta(E_n).
\end{equation}
Inserting this relation into Eq.~(\ref{eq:normalorderedexp}), the ITP can be written in terms of the operators $\hat{a}_\beta(E)$ for the incoming states. The evaluation of the ITP then amounts to calculating equilibrium averages of combinations of operators for the incoming states $\hat{a}_\beta(E)$. At zero temperature, the incoming states are all filled up to the Fermi level. (A voltage difference $V$ between the leads can be included as a time-dependent scattering phase). After some algebra, we then arrive at
\begin{equation}
  \label{eq:itpdet}
  \Pi(\tau,t_0) = \det(1 - \mathbf{Q}_{\tau,t_0})
\end{equation}
where the single-particle matrix elements of $\mathbf{Q}_{\tau,t_0}$ are given in the Supplemental Material together with the detailed derivation of Eq.~(\ref{eq:itpdet}) \footnote{See Supplemental Material at ...}. In the derivation, we concentrated on situations where all particles scattered into the right lead originate from the left lead. We consider the waiting times between particles transmitted through the scatterer above the Fermi level at $E_F=0$ of the right lead, for example using an appropriate energy filter. Equation~(\ref{eq:itpdet}) applies to a single conduction channel. For $N$ channels, the ITP reads $\Pi_N(\tau,t_0) = [\Pi(\tau,t_0)]^N$ with $N=2$ for independent spin channels.

\begin{figure*}
  \includegraphics[width=\columnwidth]{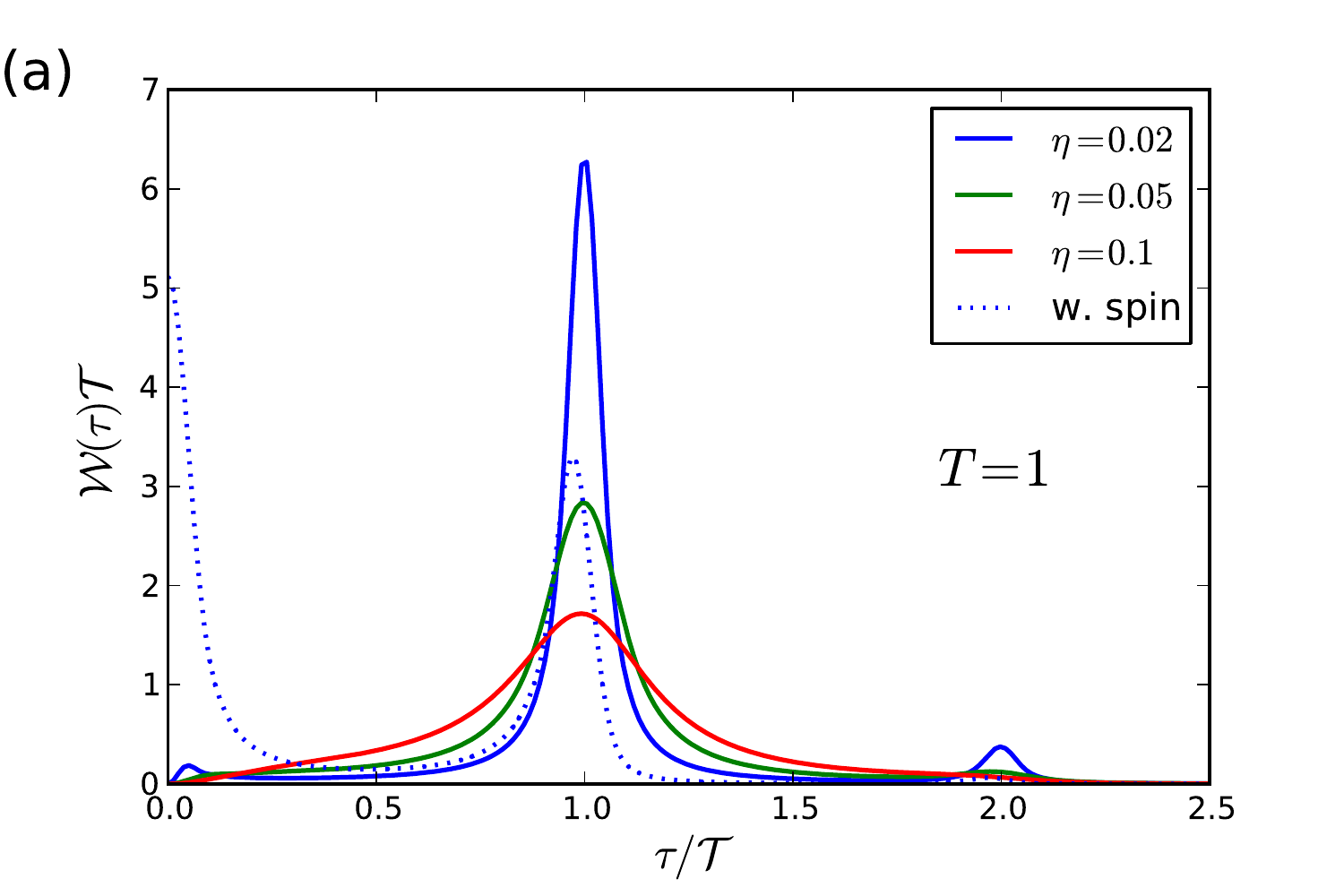}
  \includegraphics[width=\columnwidth]{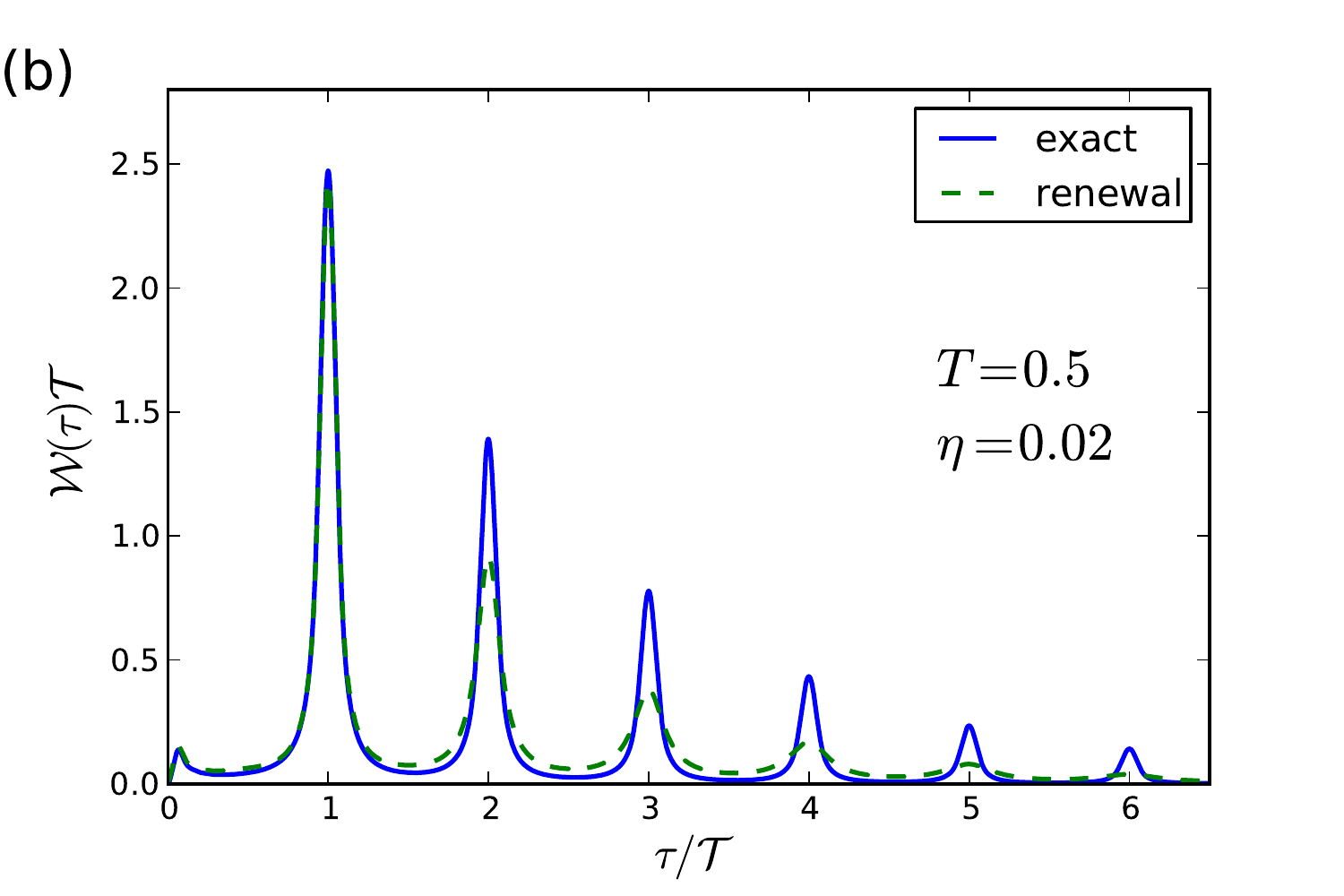}
  \caption{(color online). Waiting times between levitons. (a) WTDs for spin-less levitons with different relative widths $\eta= \tau_p / \mathcal{T}$ and the QPC fully open ($T=1$). For small widths, a clear peak in the WTD is observed at the period of the voltage pulses, $\tau=\mathcal{T}$, together with small side peaks.  As the pulses start to overlap, the peaks are smeared out. The dashed line shows results for two independent spin-channels. (b) WTD for spin-less levitons with the QPC tuned to half transmission ($T=0.5$). Cycle missing events may now occur as levitons reflect back on the QPC, giving rise to clear peaks at multiples of the period. We compare exact results to the approximation in Eq.~(\ref{eq:qpcconvolutionlaplace}) based on a renewal assumption.}
  \label{fig:wtdlevitons}
\end{figure*}

\paragraph{Driven quantum point contact.---}
We now turn to the experimentally relevant setup depicted in Fig.~\ref{fig:levitonwtdsetup}, consisting of a QPC connected to source (left) and drain (right) electrodes. We first apply a periodic voltage $V(t)$ to the left electrode and later discuss a time-dependent transmission $T(t)$. The voltage consists of a series of
Lorentzian-shaped pulses
\begin{equation}
  \label{eq:lorentzianvoltage}
  V(t) = \frac{\hbar}{e} \sum_{n=-\infty}^\infty \frac{2 \tau_p}{(t-n \mathcal{T})^2 + \tau_p^2}
\end{equation}
as illustrated in Fig.~\ref{fig:levitonwtdsetup}. The width of the pulses is $\tau_p$ and the period is $\mathcal{T}$. Remarkably, such pulses lead to clean single-electron excitations without accompanying holes as predicted in Refs.~\cite{levitov96,ivanov97,keeling06} and recently demonstrated experimentally by Dubois \emph{et al.}~\cite{dubois13nature,flindt13} and hereafter named levitons. The same outgoing state can be created by a mesoscopic capacitor with a slow linear driving protocol \cite{keeling08,hofer13,inhofer13,battista12,olkhovskaya08,haack13}.

We treat the adiabatic regime, where the time scale over which the voltage is modulated is
much longer than the time it takes an electron to pass through the scattering region.  The
Floquet scattering matrix $\mathcal{S}$ can then be related to the ``frozen'' scattering
matrix $\mathcal{S}^f(t)$ at time $t$ as
$\mathcal{S}_{\alpha\beta}(E_n,E)=\int_0^\mathcal{T}\mathrm{d} t e^{in\Omega
  t}\mathcal{S}^f_{\alpha\beta}(E,t)/\mathcal{T}$ \cite{Moskalets02,moskaletsbook11}.  The
frozen transmission amplitude is energy independent and reads $\mathcal{S}^f_{RL}(E,t) = \sqrt{T} \sin[\pi(t/\mathcal{T}+i\eta)]/\sin[\pi(t/\mathcal{T}-i\eta)]$ with $\eta = \tau_p / \mathcal{T}$, see e.~g.~Ref.~\cite{dubois13}. For the Floquet scattering amplitude, we find for $n\geq 1$
\begin{equation}
  \label{eq:lorentziansfmatrix}
  \mathcal{S}_{RL}(E_n,E) = 2 \sqrt{T} \sinh (2 \pi \eta) e^{-2 \pi \eta n}.
\end{equation}
For well-separated pulses, $\eta\ll 1$, the amplitude reduces to $\mathcal{S}_{RL}(E_n,E) \simeq 4 \pi \sqrt{T} \eta e^{-2 \pi \eta n}$ as in
Ref.~\cite{dubois13}. For pulses with a large overlap, $\eta\gg 1$, the scattering amplitude $\mathcal{S}_{RL}(E_n,E)\simeq\sqrt{T}\delta_{n,1}$ is that of a QPC with transmission $T$ and static voltage $V_{\rm dc}=\hbar\Omega/e$, see also Eq.~(\ref{eq:modulatedqpcfloquet}) below. With the Floquet scattering matrix at hand we
evaluate the ITP in Eq.~(\ref{eq:itpdet}) to find the WTD.

Figure~\ref{fig:wtdlevitons}a shows WTDs for different pulse widths with the QPC fully
open. We first consider a quantum Hall system, where the voltage pulses are applied to a
chiral edge state and the electrons can be treated as spinless (full lines). In this case,
the WTDs are suppressed to zero at $\tau=0$ due to the fermionic statistics which prevents
two electrons from occupying the same state \cite{albert12}. With sharp pulses, most
levitons are separated by one period of the driving as reflected by the large peak at
$\tau=\mathcal{T}$. However, although one excitation is created in each period, the
detection of an electron may happen in the (long) tails of the leviton, such that a period
is skipped. This gives rise to the small but visible side peaks at multiples of the period as well as the small peak just after $\tau=0$, showing that even an ideal single-electron source may suffer from cycle-missing events. In a quantum circuit, this is important for the synchronized arrival of single electrons. As the pulse width is increased, the peak in the WTD broadens as the waiting time becomes less regular. For strongly overlapping pulses, the voltage is essentially constant and we recover the results for a dc-biased QPC (not shown) \cite{albert12}.

We now turn to the experimental situation realized by Dubois \emph{et al.}~\cite{dubois13nature}, where the electronic spin is important. In this case, the WTD (dotted line) develops a large peak around $\tau=0$ corresponding to one electron in each spin channel being emitted nearly simultaneously. A second peak appears for waiting times slightly shorter than the period $\mathcal{T}$. This peak corresponds to the waiting time between the pairs of electrons that are emitted almost periodically with period $\mathcal{T}$.

Figure~\ref{fig:wtdlevitons}b shows the WTD for spinless electrons with the QPC tuned to half transmission. Levitons may now reflect back on the QPC, and cycle-missing events, in which no levitons reach the right electrode within several periods, are very likely. The cycle-missing events give rise to clear peaks at multiples of the period. The effect of the QPC can be understood in a simple picture by resolving the WTD with respect to the number of reflections that have occurred as $\mathcal{W}(\tau) = T \mathcal{W}_1^{\rm in}(\tau) + T R \mathcal{W}_2^{\rm in}(\tau) + T R^2 \mathcal{W}_3^{\rm in}(\tau) + \dots$. The reflection probability is $R=1-T$ and the $\mathcal{W}^{\rm in}_n(\tau)$'s are the distributions of waiting times between $n+1$ incoming levitons. These are related to the joint probability distributions $\mathcal{W}^{\rm in}_n(\tau_1, \dots, \tau_n)$ for $n$ successive waiting times between incoming levitons, for example $\mathcal{W}^{\rm in}_2(\tau)=\int_0^\tau \mathrm{d} t_1 \mathcal{W}_2^{\rm in}(t_1, \tau-t_1)$. Introducing the Laplace transform  $\widetilde{\mathcal{W}}(z)=\int_0^\infty \mathrm{d} \tau \mathcal{W}(\tau)e^{-z\tau}$, we have $\widetilde{\mathcal{W}}^{\rm in}_n(z)=\widetilde{\mathcal{W}}^{\rm in}_n(z,\ldots,z)$. We now make the renewal assumption that successive waiting times are uncorrelated \citep{coxbook62} such that the joint WTDs factorize as $\widetilde{\mathcal{W}}^{\rm in}_{n}(z,\dots,z) \simeq [\widetilde{\mathcal{W}}^{\rm in}_1(z)]^n$. We can then resum the geometric series $\widetilde{\mathcal{W}}(z) \simeq T \widetilde{\mathcal{W}}^{\rm in}_{1}(z)\sum_{n=0}^\infty [R\widetilde{\mathcal{W}}^{\rm in}_{1}(z)]^{n}$ as
\begin{equation}
  \label{eq:qpcconvolutionlaplace}
  \widetilde{\mathcal{W}}(z) \simeq \frac{T \widetilde{\mathcal{W}}^{\rm in}_{1}(z)}{1-R
    \widetilde{\mathcal{W}}^{\rm in}_{1}(z)}.
\end{equation}
The WTD of the incoming levitons, $\widetilde{\mathcal{W}}^{\rm in}_{1}(z)$, is the WTD at full transmission ($T=1$) shown in Fig.~\ref{fig:wtdlevitons}a.

Equation (\ref{eq:qpcconvolutionlaplace}) provides us with a direct test of the renewal
assumption of uncorrelated waiting times. Reverting it to the time domain, we can compare
Eq.~(\ref{eq:qpcconvolutionlaplace}) with the exact results in
Fig.~\ref{fig:wtdlevitons}b. The first peak around $\tau=\mathcal{T}$ is governed by the
term $T\mathcal{W}^{\rm in}_{1}(\tau)$, which does not depend on the renewal assumption,
and good agreement is found. In contrast, the following peaks are increasingly smeared out
under the renewal assumption. This demonstrates that successive waiting times are
correlated. The external driving produces a quasi periodic train of incoming
levitons. Thus, a waiting time that is shorter (longer) than the period $\mathcal{T}$ will
likely be followed by a waiting time that is longer (shorter) than the period. These
correlations, which are responsible for the sharp peaks in Fig.~\ref{fig:wtdlevitons}b,
are omitted under the renewal assumption. We note that the full counting statistics for
this problem is always binomial with success probability $T$ and, therefore, does not
distinguish between a static voltage and a series of Lorentzian pulses \cite{ivanov97}. In
contrast, the WTD captures the influence of the width of the pulses and of correlations
between single-electron emissions, which is crucial for synchronized operations of quantum
devices.

\paragraph{Time-dependent transmission.---}

We now fix the voltage $V(t)=V$ and instead modulate the transmission probability periodically in time as $T(t)=T_0[1-\epsilon \sin(\Omega t)]^2$ \citep{klich09,zhang09}. The average transmission probability is $T_{\rm av}=T_0(1+\epsilon^2/2)$ and the maximal transmission $T_{\rm max}=T_0 (1+\epsilon)^2$ must be smaller than unity. The frozen transmission amplitude from the left to the right lead reads $\mathcal{S}^f_{RL}(E,t) = \sqrt{T(t)} e^{-ieVt/\hbar}$. For the Floquet scattering matrix we find in the adiabatic limit
\begin{equation}
  \label{eq:modulatedqpcfloquet}
  \mathcal{S}_{RL}(E_n ,E) = \sqrt{T_0}[\delta_{n,p} +i\epsilon(\delta_{n,p-1}-\delta_{n,p+1})/2],
\end{equation}
assuming for the sake of simplicity that the applied voltage is a multiple of the modulation quantum, $eV=p \hbar\Omega$, where $p$ is an integer, so that the problem is
$\mathcal{T}$ periodic. (The case $qeV=p \hbar\Omega$ with $p$ and $q$ being integers can
easily be treated, although the problem becomes $q\mathcal{T}$ periodic). Apart from the central energy band (the Kronecker delta $\delta_{n,p}$) due to the voltage bias, there are two sidebands corresponding to electrons emitting ($\delta_{n,p-1}$) or absorbing ($\delta_{n,p+1}$) a modulation quantum.
\begin{figure}
  \includegraphics[width=\columnwidth]{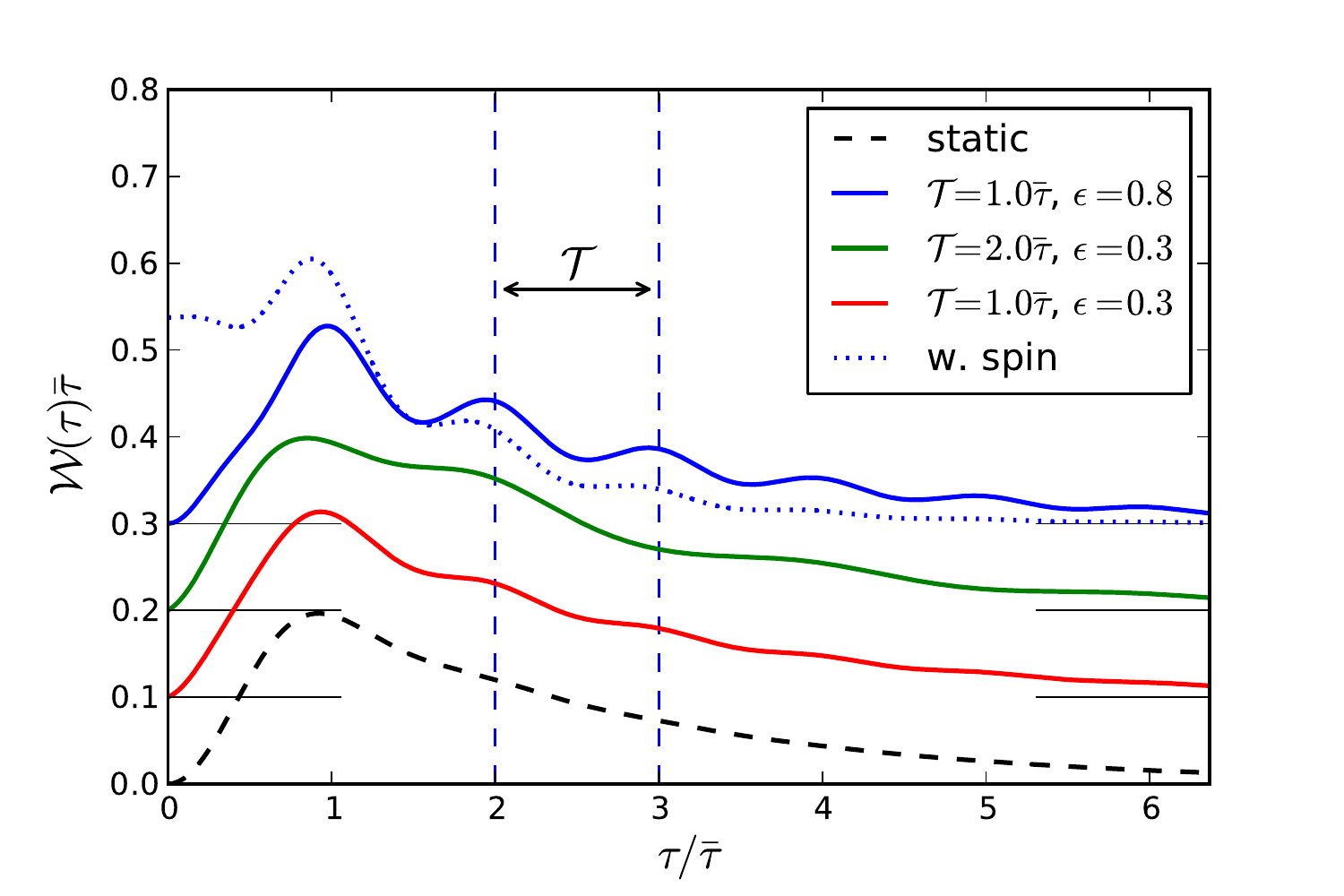}
  \caption{(color online). Periodically modulated QPC transmission. The WTDs are shifted
    vertically for clarity. Oscillations with period $\mathcal{T}$ are superimposed on the
    WTD for a static QPC with $T_\text{av}=0.4$ (dashed lines). The dotted line shows
    results for two independent spin channels}
  \label{fig:wtdmodulatedqpc}
\end{figure}

Figure~\ref{fig:wtdmodulatedqpc} shows the WTD for different modulation frequencies. Focusing first on chiral edge states, we find oscillations in the WTDs with period $\mathcal{T}$ (full lines). The oscillations are most clearly visible with large oscillation amplitudes (blue curve), demonstrating that the periodic modulation regulates the stream of incoming electrons. For comparison, we show the WTD for the static case (dashed line), where no oscillations are seen. Turning next to the situation with two independent spin channels (dotted line), the suppression of the WTD at short times is lifted, since two electrons with opposite spin may simultaneously be transmitted through the QPC.

\paragraph{Conclusions.---}
We have developed a Floquet theory of electronic WTDs in periodically driven quantum conductors. We illustrated our method by evaluating the WTDs for a driven QPC, focusing on levitons produced by Lorentzian-shaped voltage pulses as well as a periodic modulation of the QPC transmission. For both driving schemes, the WTDs provide us with a detailed characterization of the dynamic quantum conductor, beyond what can be obtained from the full counting statistics alone. Directions for future work include the WTDs for nonadiabatic driving protocols as well as investigations of correlations between electron waiting times.

\paragraph{Acknowledgements.---}
\begin{acknowledgements}
  We thank M.~Albert, P.~Devillard, G.~Haack, P.~P.~Hofer, J.~Li, M.~Moskalets, J.~R.~Ott, and P.~Roulleau for helpful discussions. In addition, we thank M.~Albert for letting us know about a recent preprint on WTDs of levitons \cite{albert14}. Our work was supported by Swiss NSF and NCCR QSIT.
\end{acknowledgements}



\section*{Supplementary material}

# Floquet Theory of Electron Waiting Times in Quantum-Coherent Conductors
## Supplemental material

David Dasenbrook, Christian Flindt, and Markus Büttiker

*Département de Physique Théorique, Université de Genève, 1211 Genève, Switzerland*

Here we derive the determinant formula for the idle time probability

$$\Pi(\tau, t_0) = \det(1 - \mathbf{Q}_{\tau,t_0}) \quad (1)$$

with the matrix elements

$$\mathbf{Q}_{\tau,t_0}(E, E') = \sum_{m=-\lfloor E/\hbar\Omega \rfloor}^{\infty} \sum_{n=-\lfloor E'/\hbar\Omega \rfloor}^{\infty} \mathcal{S}_{RL}^*(E_m, E) \mathcal{S}_{RL}(E'_n, E') \\ \times K_{\tau,t_0}(E_m, E'_n) \Theta(-E)\Theta(-E') \quad (2)$$

from the expression

$$\Pi(\tau, t_0) = \left\langle :e^{-\widehat{Q}_\tau}: \right\rangle_{t_0+\tau}. \quad (3)$$

In Eq. (2) above, the kernel reads [1, 2]

$$K_{\tau,t_0}(E, E') = 2e^{iqv_F(t_0-\tau/2)} \frac{\sin(qv_F\tau/2)}{q} \quad (4)$$

and

$$q = \frac{E - E'}{\hbar v_F}. \quad (5)$$

To begin with, we show that the average with respect to a Slater determinant of the normal-ordered exponential of a generic single-particle operator $\widehat{O}$ can be written

$$\left\langle \Psi \right| :e^{-\widehat{O}}: \left| \Psi \right\rangle = \det(\mathbf{1} - \mathbf{O}). \quad (6)$$

Here, the single-particle operator is

$$\widehat{O} = \sum_{m,n} o_{mn} \hat{a}_m^\dagger \hat{a}_n, \quad (7)$$

and the matrix $\mathbf{O}$ has the matrix elements $[\mathbf{O}]_{mn} = o_{mn}$. The operators $\hat{a}_m^\dagger$, $\hat{a}_n$ create and annihilate particles in the basis states labeled by $m$ and $n$, respectively. The many-body state is a Slater determinant of the form

$$|\Psi\rangle = \prod_{n=1}^{N} \hat{a}_n^\dagger |0\rangle. \quad (8)$$

where $N$ is the total number of particles.

To obtain Eq. (6), we first choose a basis in which the matrix $\widetilde{\mathbf{O}}$ corresponding to the operator $\widehat{O}$ is diagonal.

The determinant on the right hand side of Eq. (6) can then be written as a simple product and expanded as

$$\det(\mathbf{1} - \widetilde{\mathbf{O}}) = \prod_{n=1}^{N}(1 - \tilde{o}_{nn}) = \sum_{k=0}^{N}(-1)^k \sum_{\sigma \in \mathcal{P}_k(\mathbb{N}_N)} \prod_{i \in \sigma} \tilde{o}_{ii}, \quad (9)$$

having defined $\mathcal{P}_k(\mathbb{N}_N)$ as the set of all subsets with $k$ elements of the set of the first $N$ natural numbers, $\mathbb{N}_N$. Likewise, the left hand side of Eq. (6) can be expanded as

$$\left\langle :e^{-\widehat{O}}: \right\rangle = \sum_{k=0}^{N} \frac{(-1)^k}{k!} \left\langle :\left(\sum_{i=1}^{N} \tilde{o}_{ii} \hat{c}_i^\dagger \hat{c}_i\right)^k: \right\rangle, \quad (10)$$

where the operators $\hat{c}_i^\dagger$ and $\hat{c}_i$ are related to the operators $\hat{a}_i^\dagger$ and $\hat{a}_i$ by the change of basis.

The equality of the right hand sides of Eqs. (9) and (10) can now be established order by order using induction. We immediately see that for $k = 0$ both terms are simply 1. We now assume the equality of the terms for a given order $k$. Defining the $k$'th order term in Eq. (10) to be $\langle \hat{E}_k \rangle$ for brevity, the term of order $k+1$ reads

$$\langle \hat{E}_{k+1} \rangle = -\sum_{l=1}^{N} \frac{\tilde{o}_{ll}}{k+1} \left\langle \hat{c}_l^\dagger \hat{E}_k \hat{c}_l \right\rangle, \quad (11)$$

where we used that the operator $\hat{E}_k$ is already normal-ordered by definition. By the induction hypothesis, the expectation value of $\hat{E}_k$ with respect to a Slater determinant is given by the $k$'th order term in Eq. (9). Seeing that the operator $\hat{c}_l$ simply removes the particle labeled by $l$ from the Slater determinant, we have

$$\langle \hat{E}_{k+1} \rangle = \sum_{l=1}^{N} \frac{\tilde{o}_{ll}}{k+1}(-1)^{k+1} \sum_{\sigma \in \mathcal{P}_k(\mathbb{N}_N \setminus \{l\})} \prod_{i \in \sigma} \tilde{o}_{ii}. \quad (12)$$

Instead of summing over all $\sigma \in \mathcal{P}_k(\mathbb{N}_N \setminus \{l\})$ and multiplying by $\tilde{o}_{ll}$, we can instead sum directly over all $\sigma \in \mathcal{P}_{k+1}(\mathbb{N}_N)$. The summation over $l$ then just provides an additional factor $k+1$, so that

$$\langle \hat{E}_{k+1} \rangle = (-1)^{k+1} \sum_{\sigma \in \mathcal{P}_{k+1}(\mathbb{N}_N)} \prod_{i \in \sigma} \tilde{o}_{ii}, \quad (13)$$

which is just the $(k+1)$'st order term of Eq. (9). We thereby arrive at Eq. (6).

We now use this general result to demonstrate Eq. (1). The role of the labels $m$ and $n$, indexing the single-particle states, is played by the energy indices $E$ and

$E'$ (after a suitable regularization, see e.g. Ref. [2]) together with the lead indices $\alpha$ and $\beta$. The incoming state is a Slater determinant of all single-particle states filled up to the Fermi energy, $E_F = 0$:

$$|\Psi_{\text{in}}(t)\rangle = \prod_{\alpha=L,R} \prod_{E<0} e^{-itE} \hat{a}_\alpha^\dagger(E) |0\rangle. \quad (14)$$

We now consider the operator

$$Q_\tau = \sum_{E,E'} q_{E,E'}(\tau) \hat{b}_R^\dagger(E) \hat{b}_R(E') \quad (15)$$

counting particles in the right lead, with its matrix elements in the basis of outgoing right-moving states

$$q_{E,E'}(\tau) = \Theta(E)\Theta(E') \int_{x_0}^{x_0+v_F\tau} \phi_{R,E'}^*(x) \phi_{R,E}(x) \mathrm{d}x. \quad (16)$$

We may express it in terms of the *incoming* operators $\hat{a}_\alpha^\dagger(E)$ and $\hat{a}_\beta(E')$ using the Floquet scattering matrix,

$$\hat{b}_\alpha(E) = \sum_\beta \sum_{E_n} \mathcal{S}_{\alpha\beta}(E, E_n) \hat{a}_\beta(E_n). \quad (17)$$

The $\Theta$-functions in Eq. (16) make sure that only particles above the Fermi level are counted. Absorbing the time-dependence of the Slater determinant Eq. (14) into the operator $Q_\tau$ and taking the average at the time $t_0 + \tau$ as in Eq. (3), the matrix elements in terms of the incoming states are then given by Eq. (2). Using the result Eq. (6), we thus see that the ITP is given by Eq. (1).

---




\end{document}